\documentstyle[pre,aps,psfig,twocolumn,amsfonts,eqsecnum]{revtex}
\begin{document}
\def\u{\bbox}
\def\d{\displaystyle}
\def\mathcal#1{{\cal #1}}
\def\goldenmean{\gamma}
\def\phi{\varphi}
\def\epsilon{\varepsilon}
\def\goldenmean{\gamma}

\draft

\title{Universality for the breakup of invariant tori in Hamiltonian flows}

\author{$^1$C. Chandre, $^1$M. Govin, $^1$H. R. Jauslin, and $^2$H. Koch}

\address{$^1$Laboratoire de Physique, CNRS, Universit\'e de Bourgogne,
BP 400, F-21011 Dijon, France}
\address{$^2$Department of Mathematics, The University of Texas at Austin,
Austin, TX 78712}

\maketitle

\begin{abstract} In this article, we describe a new renormalization-group
scheme for analyzing the breakup of invariant tori for Hamiltonian systems
with two degrees of freedom.  The transformation, which acts on
Hamiltonians that are quadratic in the action variables, combines a
rescaling of phase space and a partial elimination of irrelevant
(non-resonant) frequencies. It is implemented numerically for the
case applying to golden invariant tori. We find a nontrivial fixed
point and compute the corresponding scaling and critical indices.
If one compares flows to maps in the canonical way, our results
are consistent with existing data on the breakup of golden 
invariant circles for area-preserving maps. 
\end{abstract} 
\pacs{PACS numbers: 05.45.+b, 64.60.Ak}

\section{Introduction}
In Hamiltonian systems with two (or more) degrees of freedom,
smooth invariant tori typically persist under small perturbations
\cite{kolmogorov,arnold,moser,delallave}. The most stable tori
appear to be
the ones for which the frequency ratio is a quadratic irrational,
such as the golden mean. As the strength of the perturbation
passes some critical value, these tori are observed to exhibit
self-similar scaling behavior, and then they break 
up~\cite{kadanoff,shenkerkadanoff}. To study such critical behavior,
several renormalization schemes have been proposed
\cite{escandedoveil,escande,mackay,mackayL,koch,govin,chandre}. \\
The idea is to use a transformation $\mathcal{R}$ that maps a 
Hamiltonian $H$ into a rescaled Hamiltonian $\mathcal{R}(H)$, in
such a way that irrelevant degrees of freedom are contracted.
The transformation $\mathcal{R}$ should have roughly the
following properties:
$\mathcal{R}$ has an attractive integrable fixed point $H_0$ that has a
smooth
invariant torus of a given frequency. Every Hamiltonian in its
domain of attraction $\mathcal{D}$ has 
a smooth invariant torus of the desired frequency. 
Another non-integrable fixed point $H_*$ lies on the boundary
of $\mathcal{D}$, also called critical surface, and this nontrivial
fixed point attracts the Hamiltonians on the critical surface.
Figure 1 shows schematically the expected nature of the 
renormalization flow in the space of Hamiltonians.
The existence of a nontrivial fixed point has strong implications concerning 
universal properties associated with the breakup of invariant tori.
The analysis of the renormalization for area-preserving maps~\cite{mackay}
gives support to the validity of this general picture.\\
The renormalization we define, following the scheme proposed in
Ref.~\cite{koch}, is similar
in spirit to the block spin
transformation in statistical mechanics, in the sense that it uses a process
of ``elimination and rescaling''. The elimination of irrelevant
frequencies is done by using canonical transformations as in
Kolmogorov-Arnold-Moser (KAM) theory. The frequencies we
want to eliminate are the non-slow modes
(non-resonant part of
the perturbation), i.\ e.\ the modes which only
affect the motion for a short time. The slow modes (resonant part of the
perturbation), which produce the small
denominators in KAM theory, are shifted towards the non-slow modes
by a rescaling. They can thus be eliminated by iteration. 
The rescaling is the
same as in Refs.\ \cite{koch,govin,chandre}. It includes a shift
of the 
resonances, and a rescaling of the actions and of the energy.
Here and in what follows, the word ``resonances'' refers to the 
frequency vectors ${\u \nu}_k=(p_k,q_k)$ defined by the
continued fraction approximants $p_k/q_k$ for the frequency
ratio of the invariant torus. 
The corresponding closed orbits accumulate at the invariant
torus~\cite{greene},
and are often used in numerical investigations.\\
We consider the following class of Hamiltonians
with two degrees of freedom,
quadratic in the action variables $\u{A}=(A_1,A_2)$ and described 
by three scalar functions of the angles 
${\u \varphi}=(\varphi_1,\varphi_2)\in {\Bbb T}^2$ (the two-dimensional
torus parametrized by $[0,2\pi)^2$):
\begin{eqnarray}
H({\u A},{\u \varphi})=&&
\frac{1}{2}m({\u \varphi})( {\u \Omega}\cdot{\u A})^{2} \nonumber \\
&&+\lbrack {\u \omega}_{0}
+ g({\u \varphi}){\u \Omega} \rbrack \cdot{\u A}
+ f({\u \varphi}) \label{hamiltonian} \ ,
\end{eqnarray}
where ${\u \omega_0}$ 
is the frequency vector of the considered torus, and ${\u \Omega}=(1,\alpha)$ 
is some other constant vector, not parallel to $\u{\omega}_0$.\\
This family of Hamiltonians has been investigated in Refs.\ 
\cite{govin,chandre,chandrejauslin}. In particular, a KAM theorem
was proven for this family in Ref.\ \cite{chandrejauslin} based on
Thirring's approach~\cite{thirring}, in which the KAM transformations
are constructed such that the iteration stays within the space of
Hamiltonians quadratic in the actions.
In order
to prove the existence of a torus with frequency vector 
${\u \omega}_0$ for Hamiltonian systems described 
by Eq.\ (\ref{hamiltonian}), it is not necessary to eliminate $m$,
but only $g$ and $f$:
One has to find a canonical transformation such that the equations
of motion expressed in the new coordinates show trivially
the existence of this torus. For Hamiltonians (\ref{hamiltonian}), if one
takes $g$ and $f$ equal to zero, then the resulting equations of motion
are
\begin{equation}
  \frac{d{\u A}}{dt}=-\d\frac{1}{2}\frac{\partial m}{\partial {\u \phi}}
  ({\u \Omega} \cdot {\u A})^2, \, \; \frac{d{\u \phi}}{dt}=
  m({\u \phi})({\u \Omega}\cdot {\u A}){\u \Omega}+{\u \omega}_0.
\end{equation}
Then $\u{A}=0$ defines an invariant torus, and the motion
on this torus is quasiperiodic with frequency vector
${\u \omega}_0$ 
(even if the resulting Hamiltonian is not integrable).\\
The elimination of $f$ and $g$ can be achieved with
canonical transformations with generating functions that are linear in the
action variables, and thus map the family of Hamiltonians
(\ref{hamiltonian}) into itself.
This is very convenient numerically, as one only works with three scalar 
functions $m$, $g$ and $f$. The only approximation involved in our 
numerical implementation
of the transformation is a truncation of the Fourier series of these functions,
e.\ g.\ we approximate
\begin{equation}
f(\u{\phi})=\sum_{\nu\in {\Bbb Z}^2} f_{\nu} e^{i\nu\cdot\phi},
\end{equation}
by
\begin{equation}
f^{[\leq L]}(\u{\phi})=\sum_{\nu\in\mathcal{C}_L} f_{\nu} e^{i\nu\cdot\phi},
\end{equation}
where $\mathcal{C}_L=\{\u{\nu}\in {\Bbb Z}^2\left| |\nu_1|\leq L, \, 
|\nu_2|\leq L\right.\}$.\\
In this paper, we focus on the frequency vector 
${\u \omega}_0=(1/\gamma,-1)$,
associated with the golden mean $\gamma=(1+\sqrt{5})/2$.
We choose the set of frequencies $I^-$ describing the non-slow modes
as the union of the two quadrants in the plane 
$(\nu_1, \nu_2)$ that contain the linear span of
${\u \omega}_0$:
\begin{equation}
I^-=\{ \u{\nu}\in {\Bbb Z}^2 \vert \nu_1\nu_2\leq 0\}.
\end{equation}
The set $I^-$ is depicted in Fig.\ 2.\\
This ``frequency cutoff'' restricts to Fourier modes (the
non-slow modes) that can be eliminated in one renormalization step,
without running into small denominator problems. As is common
with cutoffs, there is not a single ``natural'' choice. 
Other possible choices of $I^-$ will be mentioned later.

\section{KAM-Renormalization-Group transformation}
\label{sect:kamrg}
The renormalization scheme described in this section
is for a torus of frequency vector $\u{\omega}_0=(1/\goldenmean,-1)$ where
$\goldenmean=(1+\sqrt{5})/2$. It is straightforward to adapt it to
other reduced quadratic irrationals~\cite{lang}.\\
Our transformation is composed of four steps:\\
(1) A canonical change of coordinates, which acts on the Fourier
mode $e^{i\nu\cdot\varphi}$ for a resonance ${\u \nu}={\u \nu}_k$
by a shift ${\u \nu}_k \mapsto {\u \nu}_{k-2}$. The resonances
for $k<2$ will be eliminated in step 4, together with
all the other frequency vectors in $I^-$.\\
(2) A generalized canonical change of coordinates which corresponds
to a rescaling of the action variables ${\u A}$.\\  
(3) A normalization corresponding to a rescaling of the energy
(or time).\\
The composition of the three steps described so far defines a map
$(m,g,f,\alpha)\mapsto (m',g',f',\alpha')$.\\
(4) A KAM transformation (canonical change of variables) 
that eliminates all non-slow modes from $g'$ and $f'$, i.\ e.\
the new functions $f''$ and $g''$ are in the range of ${\Bbb I}^-$,
where ${\Bbb I}^-$ is a projection operator acting
on a scalar function $f$ of the angles as 
\begin{equation}
{\Bbb I}^-f({\u \varphi})=\sum_{\nu\in I^-}f_\nu e^{i\nu\cdot\varphi}.
\end{equation}
Concerning step 4 we note that, for certain purposes
it can be desirable to eliminate the non-slow modes from $m'$ as well. 
But such a step generates terms of arbitrary order in the actions,
which drastically complicates the analysis~\cite{kochwittwer}.\\
We now give a more detailed description of the four steps that
define the transformation. The first step is
motivated by the observation that periodic orbits for the
resonant frequencies $\{ {\u\nu}_k\}$ accumulate geometrically at the
(critical)
${\u \omega}_0$--torus. This suggests that the appropriate
scaling (in the angle variables) is related to a shift in the
sequence of resonances. In order to implement the shift 
${\u \nu}_k\mapsto {\u \nu}_{k-2}$ mentioned in step 1, 
we use the fact that the continued fraction approximants for
the golden mean $\gamma$ are $\gamma_k=F_{k+1}/F_k$, where
$F_k$ is the $k$-th Fibonacci number. These numbers can be 
defined recursively by the equation
\begin{equation}
\d\left( \begin{array}{c} F_{k+1} \\ F_k \end{array} \right)=
N\d\left( \begin{array}{c} F_k \\ F_{k-1} \end{array} \right),
\; \mbox{ with } N=\d\left( \begin{array}{cc} 1 & 1 \\
1 & 0 \end{array} \right),
\end{equation}
starting with $F_0=0$ and $F_1=1$. In other words, if we define
${\u\nu}_0=(0,1)$ then ${\u\nu}_k=N{\u\nu}_{k-1}$ for $k\geq1$.
Thus the desired shift ${\u \nu}_k\mapsto{\u \nu}_{k-2}$ is
induced by the linear transformation $N^{-2}$, and the
canonical transformation we are looking for is
$(\u{A},\u{\varphi})\mapsto (N^{-2}\u{A},N^2\u{\varphi})$.\\ 
As the renormalization changes the scale,
some of the slow modes become non-slow modes [e.\ g., 
$(1,1)$ is mapped into $(0,1)$ which is
an element of $I^-$].
We remark that there is no intrinsic sharp difference between the
slow and non-slow modes. The boundary can be set at different places.
We have chosen to include the coefficients $(0,1)$ and $(1,0)$ among
the non-slow modes ($I^-$, to be eliminated by the KAM transformation),
and $(1,1)$ and $(2,1)$ among the slow modes. But conceptually
and practically there would be no difficulty to include e.\ g.\
$(1,1)$ and $(2,1)$ (or any fixed finite number of resonances)
among the non-slow modes. More generally, other choices in the
splitting of $\{ e^{i\nu\cdot\varphi}\}$ into slow and non-slow modes
should lead to the same results provided e.\ g.\ that the ratio
$|{\u \nu}|/|{\u \omega}_0\cdot{\u \nu}|$ is bounded on $I^-$, and
that $N^{-k}$ contracts vectors ${\u \nu}$ in the complement of
$I^-$ for some fixed $k>0$~\cite{koch}.\\
The linear shift of the resonances multiplies ${\u \omega}_0$ by 
$\goldenmean^{-2}$ ($\u{\omega}_0$ is an eigenvector of $N$);
therefore we rescale the energy by a factor $\goldenmean^2$ in order to keep
the frequency fixed at ${\u \omega}_0$. A consequence of the shift of
the resonances is that $\u{\Omega}=(1,\alpha)$ 
is changed into $\u{\Omega}'=(1,\alpha')$,
where $\alpha'=(\alpha+1)/(\alpha+2)$.\\
Then we perform a rescaling of the action variables: we change the  
Hamiltonian $H$ into 
$$
\hat{H}({\u A},{\u \varphi})=
\lambda H\left(\d\frac{{\u A}}{\lambda},{\u \varphi}\right),
$$ 
with $\lambda=\lambda(H)$
such that the mean value of $m'$ defined as
\begin{equation}
\langle m' \rangle=\int_{{\Bbb T}^2} \frac{d^2{\u\varphi}}{(2\pi)^2}
m'({\u\varphi}),
\end{equation}
is equal to $1$.
Since the rescaling of energy and the shift $N^2$ transform the 
quadratic term of the Hamiltonian into 
$\goldenmean^2(2+\alpha)^2m(\u{\phi})(\u{\Omega}'\cdot \u{A})^2/2$, 
this condition leads to
$\lambda=\goldenmean^2(2+\alpha)^2\langle m \rangle$.\\
In summary, the first three steps of our renormalization
transformation rescale $m$, $g$, $f$ and $\u{\Omega}=(1,\alpha)$ into
\begin{eqnarray}
     && m'(\u{\phi})=\d \frac{m\left(N^{-2}\u{\phi}\right)}
                     {\langle m \rangle},\\
     && g'(\u{\phi})=\goldenmean^2(2+\alpha) g\left(N^{-2}\u{\phi}
                      \right),\\
     && f'(\u{\phi})=\goldenmean^4(2+\alpha)^2 \langle m \rangle
                     f\left(N^{-2}\u{\phi}\right),\\
     && \alpha'=\frac{1+\alpha}{2+\alpha}.\label{eqn:alpha}
\end{eqnarray}
We remark that the map ${\u \Omega}\mapsto{\u \Omega}'$ given 
by Eq.\ (\ref{eqn:alpha}) has an attractive fixed point
${\u \Omega}_*=(1,\gamma^{-1})$, which is an eigenvector
of $N^2$, with eigenvalue $\gamma^2>1$.\\
The fourth step is carried out via an iterative procedure,
similar to a Newton algorithm.
We start with $H_0=H'$. To simplify the description
of the iteration step $H_n\mapsto H_{n+1}$, consider first
the case where $g_n$ and $f_n$ depend on a (small) parameter 
$\varepsilon_n$, in such a way that ${\Bbb I}^-g_n$ and ${\Bbb I}^-f_n$
are of order $\mathcal{O}(\varepsilon_n)$. The idea is to eliminate
the non-slow modes of $g_n$ and $f_n$ to first order in
$\varepsilon_n$, at the expense of adding terms that are of order
$\mathcal{O}(\varepsilon_n)$ in the slow modes and of order
$\mathcal{O}(\varepsilon_n^2)$ in the non-slow modes.\\
We will work with Lie transformations 
$ U_n: (\u{\varphi}_{n-1},\u{A}_{n-1})\mapsto(\u{\varphi}_{n},
\u{A}_{n})$
generated by a function $S_n$ linear in the action variables, of the form
\begin{equation}
\label{eqn:S}
S_n(\u{A},\u{\varphi})=Y_n(\u{\varphi})\u{\Omega}'\cdot\u{A}
+Z_n(\u{\varphi})+a_n\u{\Omega}'\cdot\u{\varphi}\ ,
\end{equation}
characterized by two scalar functions $Y_n$, $Z_n$, and a constant $a_n$.
The expression of the Hamiltonian in the new variables is obtained
by the following equation~\cite{deprit,benettin}:
\begin{eqnarray}
H_{n+1}&=&H_n\circ U_n=
e^{+\hat{S_n}}
H_n \nonumber\\
&\equiv& H_n+\lbrace S_n,H_n \rbrace
+\frac{1}{2!}\lbrace S_n,\lbrace S_n,H_n\rbrace\rbrace+\cdots .
\label{eqn:exponential}
\end{eqnarray}
A consequence of the linearity of $S_n$ in $\u{A}$ is that 
the Hamiltonian $H_{n+1}$ is again quadratic in the actions, and of the form 
\begin{eqnarray}
  H_{n+1}({\u A},{\u \varphi})=&&
                \frac{1}{2}m_{n+1}({\u \varphi})( {\u \Omega}'\cdot{\u A})^{2}
		\nonumber \\
              &&  +\lbrack {\u \omega}_0
                + g_{n+1}({\u \varphi}){\u \Omega}' \rbrack \cdot{\u A}
                + f_{n+1}({\u \varphi}).\label{image}
\end{eqnarray}
We notice that the vector $\u{\Omega}'$ remains unchanged from one step
to the next.
The functions $Y_n$, $Z_n$, and the constant $a_n$ are chosen in such a
way that ${\Bbb I}^-g_{n+1}$ and ${\Bbb I}^-f_{n+1}$ vanish to order
$\mathcal{O}(\varepsilon_n)$; see Appendix \ref{appendix} for details.
Consequently,
${\Bbb I}^-g_{n+1}$ and ${\Bbb I}^-f_{n+1}$ are of order 
$\mathcal{O}(\varepsilon_0^{2^n})$, where $\varepsilon_0$ denotes the
order of ${\Bbb I}^-g'$ and ${\Bbb I}^-f'$.
If $\varepsilon_0$ is small, an iteration of this procedure should define
a canonical transformation 
$U_H=U_1\circ U_2\circ \cdots \circ U_n \circ \cdots$
such that the Hamiltonian expressed in these new variables has only slow
modes in $g$ and $f$. In our numerical implementation this is indeed
the case, even without small parameters.\\
In summary, our renormalization-group (RG) transformation 
acts as follows: First, some of the slow modes (resonant 
part of the perturbation) are turned into non-slow modes by a frequency
shift and a rescaling. Then, a KAM-type iteration
eliminates these non-slow (non-resonant) modes, while producing some new
slow modes.


\section{Determination of the critical coupling; fixed point of the KAM-RG 
transformation}
\label{sect:result}

We start with the same initial Hamiltonian as in
Refs.\ \cite{escande,govin,chandre} 
\begin{equation}
\label{hamiltonieninit}
H({\u A},{\u \varphi})=\frac{1}{2}({\u \Omega}\cdot{\u A})^{2}
+{\u \omega}_{0}\cdot{\u A}+\varepsilon f({\u \varphi}) \ ,
\end{equation}
where ${\u \Omega}=(1,0)$, ${\u \omega}_0
=(1/\goldenmean,-1)$,
$\goldenmean=(1+\sqrt{5})/2$, and a perturbation
\begin{equation}
f({\u \varphi})=\cos({\u \nu}_1\cdot{\u \varphi})
+ \cos({\u \nu}_2\cdot{\u \varphi}), 
\end{equation}
where ${\u \nu}_1=(1,0)$ and ${\u \nu}_2=(1,1)$.
We represent all the
functions by their Fourier series truncated by retaining only the coefficients in
the square $\mathcal{C}_L$ which contains $(2L+1)^2$ Fourier coefficients.
For fixed $L$, we take successively larger couplings
$\varepsilon$ and determine whether the KAM-RG iteration converges to a 
Hamiltonian with 
$f=0,\,  g=0$ (trivial fixed point), or whether it diverges 
($f,g\rightarrow\infty$).
By a bisection procedure, we determine the critical
coupling $\varepsilon_c(L)$. We repeat the calculation with larger numbers
of Fourier coefficients, to obtain a more accurate approximation.
Table I lists some values of $\varepsilon_c(L)$. For 
instance, for $L=10$, the result is given with 7 digits; as a 
comparison, the method developed in Refs.\ \cite{govin,chandre}
yields 4 significant digits (in order to obtain 7 significant digits with
this method
one needs to calculate $\varepsilon_c(L)$ up to $L=34$). 
Moreover, we observe the disappearance of the oscillations of Fig.\ 1 of
Ref.\ \cite{govin}, that is to say, the present method
converges much faster than the one of Refs.\ \cite{govin,chandre}.\\

By iterating the transformation starting from a point on the critical
surface, we observe that the process converges to a nontrivial fixed point
$H_*$ (or more generally to a nontrivial fixed set related to this
nontrivial fixed point by symmetries~\cite{chandre,mackay2}),
which we characterize by the Fourier coefficients of the three functions
 $m_*,g_*,f_*$ and ${\u \Omega}_*=(1,\goldenmean^{-1})$.\\
Figures 3 and 4 show the weight of the Fourier
coefficients of $m_*$. 
We observe that
these coefficients decrease only slowly in the direction of
${\u \omega}_0$, and in particular, along the rescaled resonances 
$N^{-k}{\u \nu}_1$, $k\geq 0$, which indicates
that $m_*$ is not analytic. This is due to the fact that our
transformation does not eliminate the non-slow modes of $m$ (for
the reasons mentioned earlier).\\
Figures 5 and 6 show the weight of the
Fourier coefficients for the functions
$g_*$ and $f_*$, respectively. These coefficients seem to decrease
exponentially, which indicates that $g_*$ and $f_*$ are real 
analytic. Notice also that, by construction, $g_*$ and $f_*$
have only slow modes.\\
At the fixed point $H_*$, we compute the relevant eigenvalues
(critical exponents) for the linearized KAM-RG transformation.
Since the breakup of invariant tori is observed in one-parameter
families of Hamiltonians, indicating that the critical surface is
of codimension one, one expects to find a simple eigenvalue 
$\delta>1$, and no other spectrum outside the open unit disk. 
This is precisely what we find numerically.
Table I lists the values of $\delta$ as a function of $L$.
We obtain $\delta\in [2.650175, 2.650234]$,
which is in agreement with the value obtained by MacKay for area-preserving
maps\cite{mackay} ($\delta=2.650221$).\\
If the renormalization-group picture is correct, then all the relevant
informations about critical tori with frequency vector 
${\u \omega}_0=(1/\gamma,-1)$ is contained in the nontrivial
fixed point $H_*$.
In particular, $H_*$ determines the observed critical
scaling of phase space~\cite{kadanoff,shenkerkadanoff,escande}.
For the scaling factor $\lambda_*=\lambda(H_*)$,
we obtain numerically $\lambda_*\in [18.827910,18.828203]$;
the value obtained for
area-preserving maps is $\lambda_*=18.828171$ (given in Refs.\ 
\cite{kadanoff,shenkerkadanoff,mackay}).

\section{Conclusion}
We have shown that in the renormalization-group approach to
critical invariant tori,
a partial elimination of non-resonant frequencies
leads to substantial qualitative and quantitative improvements.
Compared with previous schemes that have been studied numerically,
our new KAM-RG transformation yields more accurate results and a
better defined domain of attraction for the nontrivial fixed point.
Conceptually it is close
to the type of transformations used for the study of 
critical phenomena in statistical mechanics.\\
We have implemented our transformation for the case of
tori with frequency
vector $\u{\omega}_0=(1/\goldenmean,-1)$, where $\gamma$ is the golden
mean. 
The extension to other reduced
quadratic irrationals should be straightforward and yields similar
results.
It also seems possible to extend our transformation to systems
with three~\cite{mackaymeissstark} and more degrees of
freedom~\cite{koch}, where very little is known about the behavior
of invariant tori under perturbations that are not necessarily 
small~\cite{maohelleman,artuso}. 

\section*{acknowledgments}

We thank G. Benfatto, G. Gallavotti, R. S. MacKay,
and P. Wittwer for helpful discussions.
H. Koch acknowledges support by the National Science Foundation
under Grant No.\ DMS-9705095.
Support from the EC contract No.\ ERBCHRXCT94-0460 for the project
``Stability and universality in classical mechanics''
and from  the Conseil R\'egional de Bourgogne is also acknowledged.

\appendix
\section{One step of the KAM iteration}
\label{appendix}
In this appendix, we use the notation ${\u \partial}f=
\partial f/\partial {\u \varphi}$.\\
We consider a Hamiltonian of the form (\ref{hamiltonian}), 
with $g=(1-{\Bbb I}^-)G+\varepsilon {\Bbb I}^- G$ and
$f=(1-{\Bbb I}^-)F+\varepsilon {\Bbb I}^-F$. The parameter
$\varepsilon$ is introduced for bookkeeping purposes only;
it will be set to one at the end. Our goal is to find a canonical 
transformation $U$ such that the functions $f'$ and $g'$ for the
Hamiltonian $H'=H\circ U$ have only slow modes, up to terms of
order $\mathcal{O}(\varepsilon^2)$.
We perform a Lie transformation defined in Sec.\ \ref{sect:kamrg}. The
expression of the Hamiltonian in the new variables is given by Eq.\ 
(\ref{eqn:exponential}):
\begin{eqnarray}
  H'({\u A}',{\u \varphi}')=&&
                \frac{1}{2}m'({\u \varphi}')( {\u \Omega}\cdot{\u A}')^{2}
		\nonumber \\
	&&+\lbrack {\u \omega}_0
                + g'({\u \varphi}'){\u \Omega} \rbrack \cdot{\u A}'
                + f'({\u \varphi}'),\label{eqn:im}
\end{eqnarray}
where
\begin{eqnarray}
\label{eqn:projg}
 g'&=&g+\u{\omega}_0\cdot\u{\partial} Y
                  +m(\u{\Omega}\cdot\u{\partial} Z
                    +a\Omega^2) +\mathcal{O}(\epsilon^2),\\
\label{eqn:projf} 
 f'&=&f+\u{\omega}_0\cdot\u{\partial} Z
 + a \u{\omega}_0\cdot\u{\Omega}+\mathcal{O}(\epsilon^2),
\end{eqnarray}
with $Y$, $Z$ and $a$ of order $\mathcal{O}(\varepsilon)$ to be
determined.
The generating function $S$ given by Eq.\ (\ref{eqn:S}) is determined by
the projection of Eqs.\ (\ref{eqn:projg})-(\ref{eqn:projf}) on the space
of non-slow modes.
We recall that the condition is that ${\Bbb I}^-f'$ and ${\Bbb I}^-g'$ are
of order $\mathcal{O}(\varepsilon^2)$.
This leads to the following equations:
\begin{eqnarray}
&& {\Bbb I}^-f+\u{\omega}_0\cdot\u{\partial}Z=const,\\
&& {\Bbb I}^-g+\u{\omega}_0\cdot\u{\partial}Y
+{\Bbb I}^-\left(m\u{\Omega}\cdot\u{\partial}Z\right)
+{\Bbb I}^-m a\Omega^2 =0. \label{eqn:tcan2}
\end{eqnarray}
Equation (\ref{eqn:tcan2}) determines $a$:
\begin{equation}
a=-\frac{\langle g\rangle+\langle m\u{\Omega}\cdot\u{\partial}Z\rangle}
{\Omega^2\langle m \rangle},
\end{equation}
Moreover the functions $Y$ and $Z$ have only non-slow modes, and
are given by the following series:
\begin{equation}
Z(\u{\varphi})=\sum_{\nu\in I^{-*}} 
\frac{i}{\u{\omega}_{0}\cdot\u{\nu}}f_{\nu} e^{i\nu\cdot\phi},
\end{equation}
\begin{eqnarray}
Y(\u{\varphi})=\sum_{\nu\in I^{-*}} \frac{i}{\u{\omega}_{0}\cdot\u{\nu}}
&& \left(g_{\nu}+(m\u{\Omega}\cdot\u{\partial}Z)_{\nu} \right. \nonumber \\
&&\left. +m_{\nu}a \Omega^2\right)e^{i\nu\cdot\phi},
\end{eqnarray}
where $I^{-*}=I^-\setminus \{ {\u 0} \}$.
The transformed Hamiltonian (\ref{eqn:im}) is constructed by defining
$H^{(0)}=H$ and $H^{(i)}$ for $i=1,2,\ldots$ by the recursive relation
\begin{eqnarray}
H^{(i+1)}(\u{A},\u{\varphi})&=&
\lbrace S(\u{A},\u{\varphi}),H^{(i)}(\u{A},\u{\varphi}) \rbrace \nonumber\\
&=&
\frac{1}{2}m^{(i+1)}(\u{\varphi})(\u{\Omega}\cdot\u{A})^{2} \nonumber \\
&& +g^{(i+1)}(\u{\varphi})\u{\Omega}\cdot\u{A}+f^{(i+1)}(\u{\varphi}),
\end{eqnarray}
which leads to
\begin{equation}
H'=\sum_{i=0}^\infty \frac{H^{(i)}}{i!}.
\end{equation}
This can be expressed in terms of the image of the three scalar functions 
$(m,g,f)$ given by the following equations:
\begin{eqnarray}
&& (m',g',f')=\left(\sum_{i=0}^{\infty} \frac{m^{(i)}}{i!},
\sum_{i=0}^{\infty} \frac{g^{(i)}}{i!},\sum_{i=0}^{\infty} \frac{f^{(i)}}{i!}
\right),\\
&& (m^{(0)},g^{(0)},f^{(0)})=(m,g,f),\\
&& m^{(1)}=2m\u{\Omega}\cdot\u{\partial}Y
           -Y\u{\Omega}\cdot\u{\partial}m,\\
&& g^{(1)}=g\u{\Omega}\cdot\u{\partial}Y
           -Y\u{\Omega}\cdot\u{\partial}g \nonumber \\
           && \qquad +m\u{\Omega}\cdot\u{\partial}Z+m a\Omega^2
           +\u{\omega}_0\cdot\u{\partial}Y,\\
&& f^{(1)}=-Y\u{\Omega}\cdot\u{\partial}f
           +g\u{\Omega}\cdot\u{\partial}Z+g a\Omega^2 
           +\u{\omega}_0\cdot\u{\partial} Z, \\
&& m^{(i+1)}=2m^{(i)}\u{\Omega}\cdot\u{\partial}Y
             -Y\u{\Omega}\cdot\u{\partial}m^{(i)}, \\
&& g^{(i+1)}=g^{(i)}\u{\Omega}\cdot\u{\partial}Y
             -Y\u{\Omega}\cdot\u{\partial}g^{(i)} \nonumber \\
             && \qquad +m^{(i)}\u{\Omega}\cdot\u{\partial}Z+m^{(i)}a\Omega^2,\\
&& f^{(i+1)}=-Y\u{\Omega}\cdot\u{\partial}f^{(i)}
             +g^{(i)}\u{\Omega}\cdot\u{\partial}Z+g^{(i)}a\Omega^2,
\end{eqnarray}
for $i\geq 1$.
\begin{table}
\label{table:par}
\caption{Parameters associated with the breakup of noble tori.}
\begin{tabular}{c c c c}
L & Critical value $\varepsilon_c(L)$ & Area multiplier & Unstable eigenvalue\\
\hline
 & 0.027590 & 18.828171 & 2.650221 \\
\hline
3 & 0.027625 & 18.842654 & 2.649660 \\
5 & 0.027579 & 18.823481 & 2.652722 \\
8 & 0.027588 & 18.825034 & 2.649653 \\
12 & 0.027590 & 18.829142 & 2.650484 \\
13 & 0.027590 & 18.827464 & 2.650082 \\
20 & 0.027590 & 18.828203 & 2.650234 \\
21 & 0.027590 & 18.827772 & 2.650151 \\
33 & 0.027590 & 18.828177 & 2.650226 \\
34 & 0.027590 & 18.827910 & 2.650175
\end{tabular}
\end{table}

\begin{figure}
\large
\unitlength=1cm
\centerline{
\begin{picture}(15,10)
\put(1.5,0){\psfig{figure=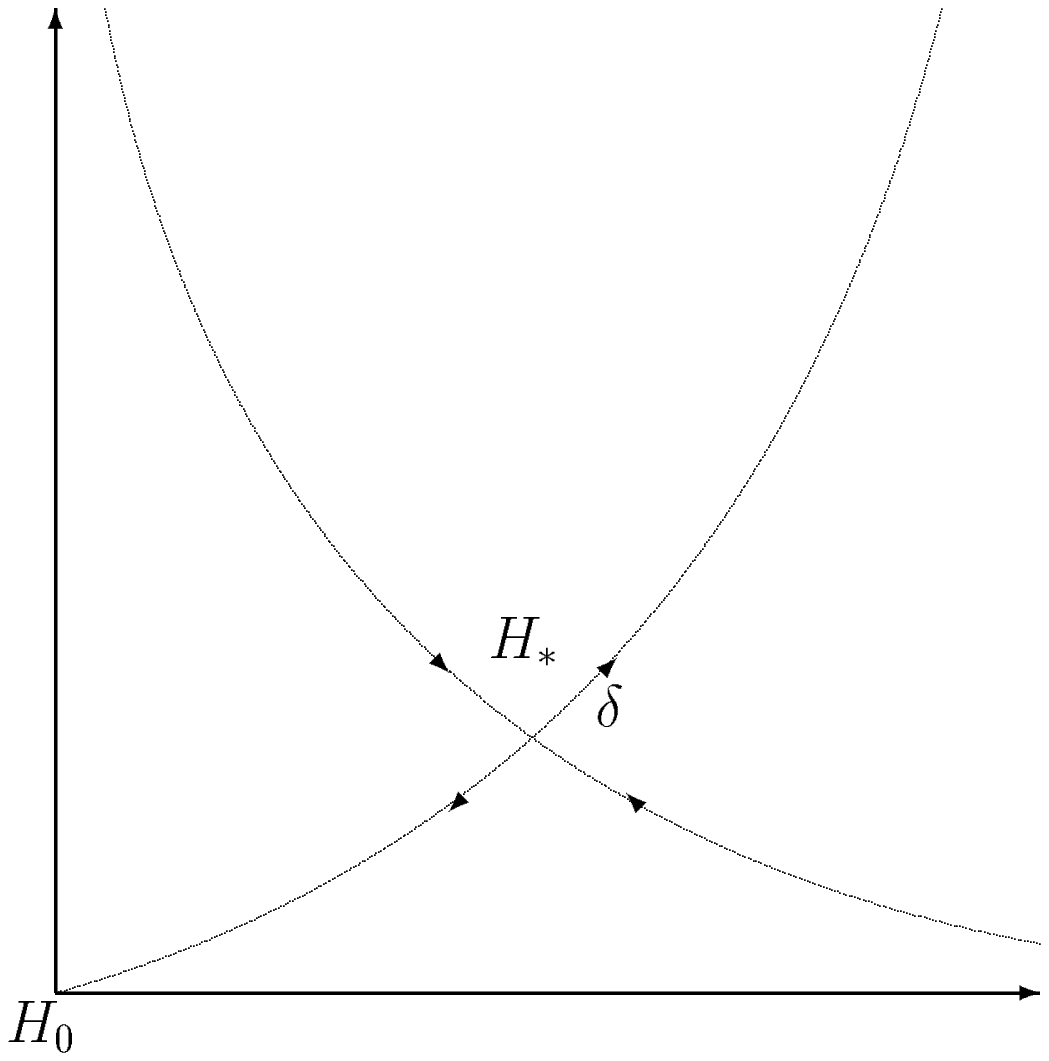,height=15cm,width=10cm}}
\normalsize 
\put(3,2){FIG.\ 1. Renormalization flow in the space of Hamiltonians}
\put(3,1.6){(\ref{hamiltonian}) associated with the universal class of 
${\u \omega}_0=(1/\gamma,-1)$.}
\end{picture}}
\end{figure}

\begin{figure}
\large
\unitlength=1cm
\centerline{
\begin{picture}(15,10)
\put(1.5,0){\psfig{figure=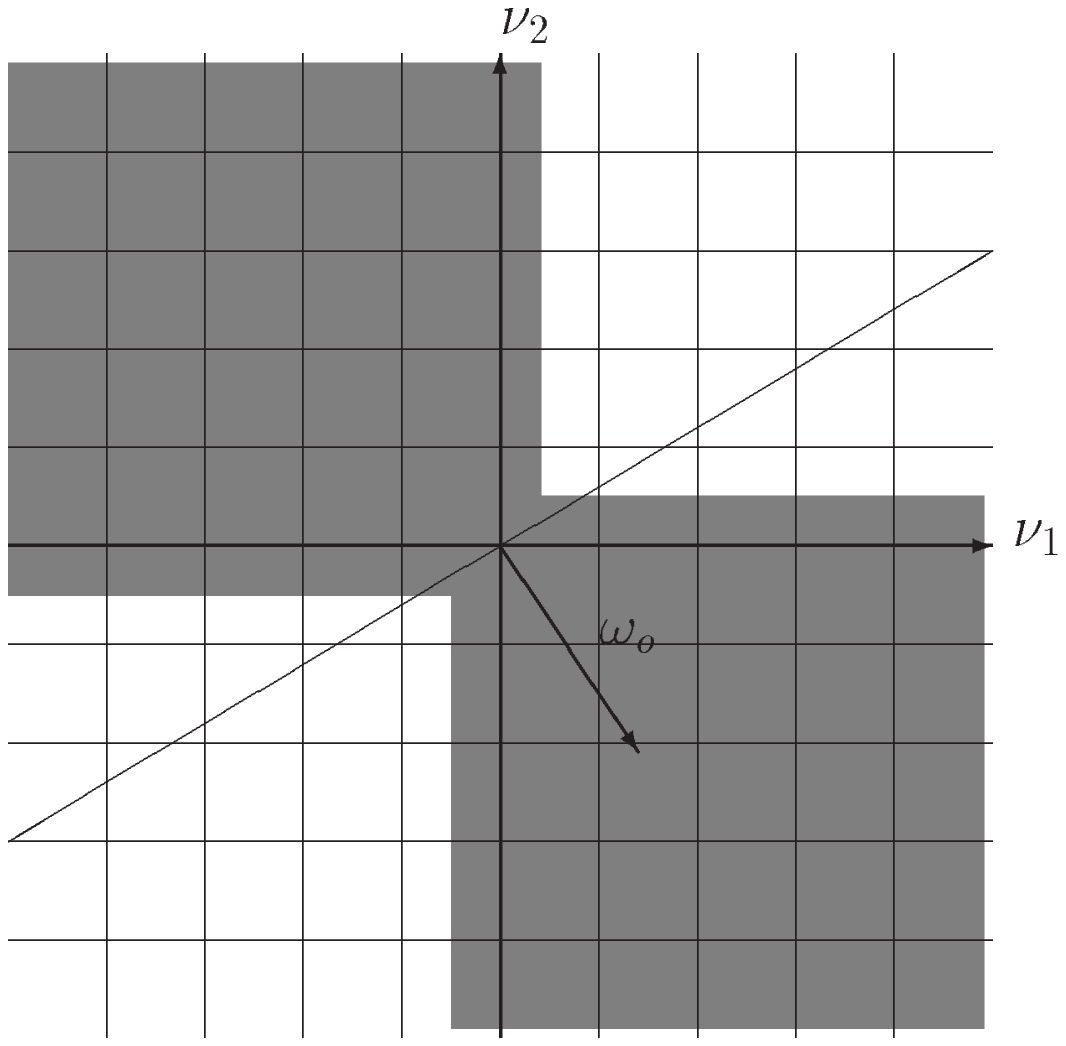,height=15cm,width=10cm}}
\normalsize 
\put(3,2){FIG.\ 2. Non-slow modes (in the grey part) and slow modes}
\put(3,1.6){(in the white part) associated with ${\u \omega}_0=(1/\gamma,-1)$.}
\end{picture}}
\end{figure}

\begin{figure}
\large
\unitlength=1cm
\centerline{
\begin{picture}(15,10)
\put(1.5,0){\psfig{figure=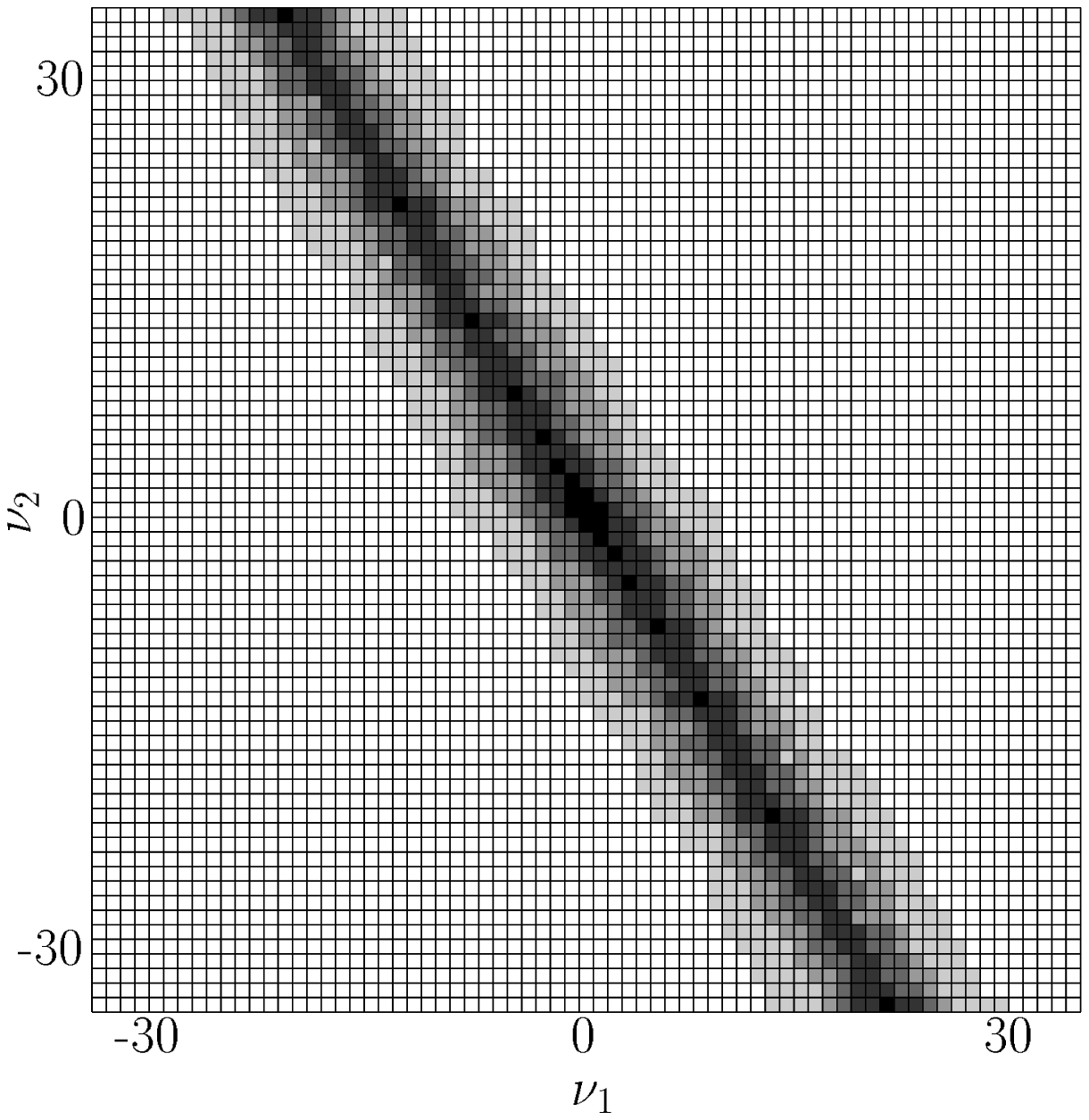,height=15cm,width=10cm}}
\normalsize 
\put(3,2){
FIG.\ 3. Weight of the Fourier coefficients of $m_*$:}
\put(3,1.6){White: $\, <10^{-10}$, grey levels: $[10^{-10},10^{-7}]$,}
\put(3,1.2){$[10^{-7},10^{-5}]$,$[10^{-5},10^{-3}]$,
$[10^{-3},5.10^{-2}]$, black: $>5.10^{-2}$.}
\end{picture}}
\end{figure}

\begin{figure}
\large
\unitlength=1cm
\centerline{
\begin{picture}(15,10)
\put(1.5,0){\psfig{figure=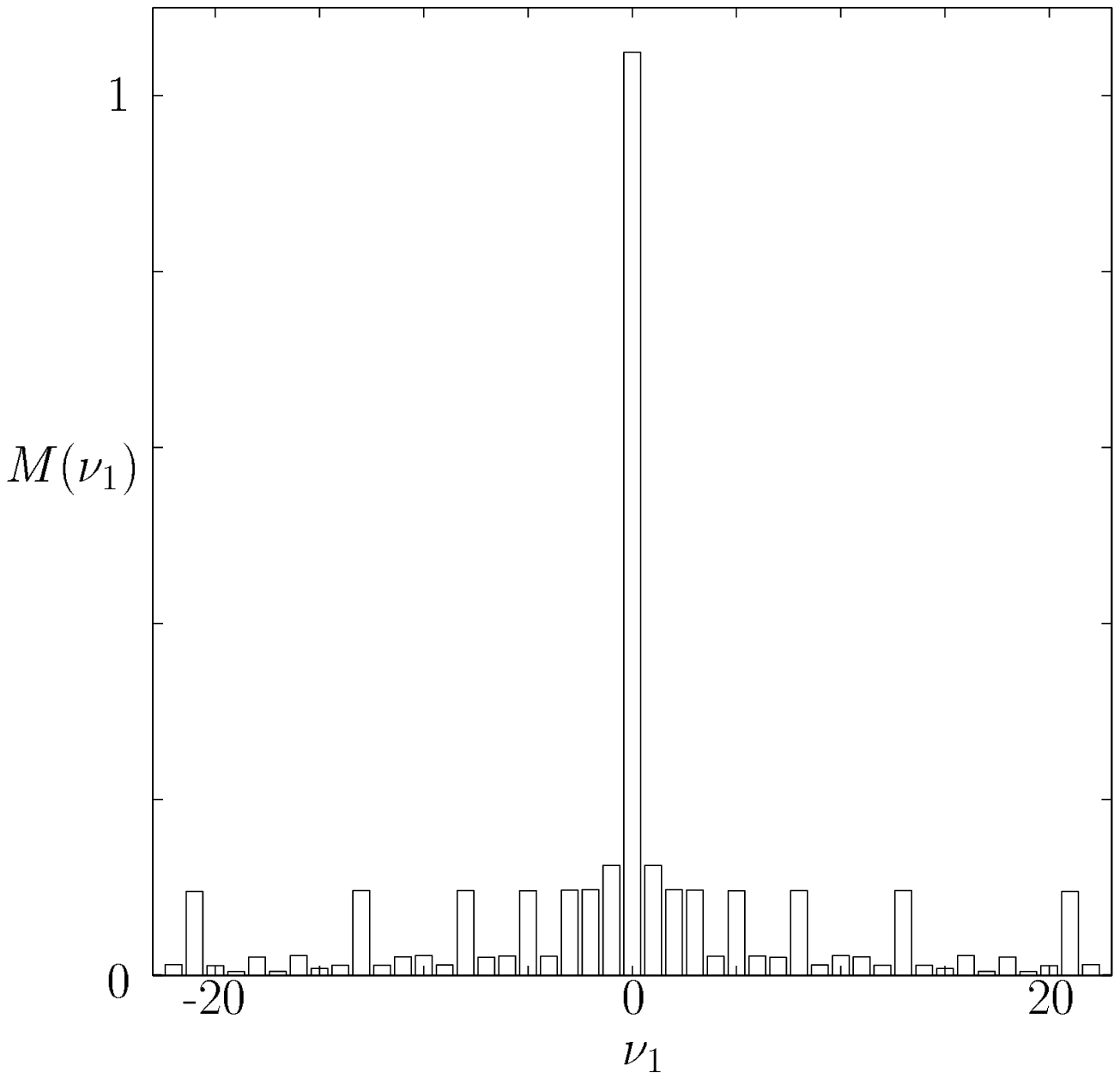,height=15cm,width=10cm}}
\normalsize 
\put(3,2){
FIG.\ 4. Weight of the Fourier coefficients of $m_*$:}
\put(3,1.6){$M(\nu_1)=\max_{\nu_2} |m_{*\nu}| $.}
\end{picture}}
\end{figure}

\begin{figure}
\large
\unitlength=1cm
\centerline{
\begin{picture}(15,10)
\put(1.5,0){\psfig{figure=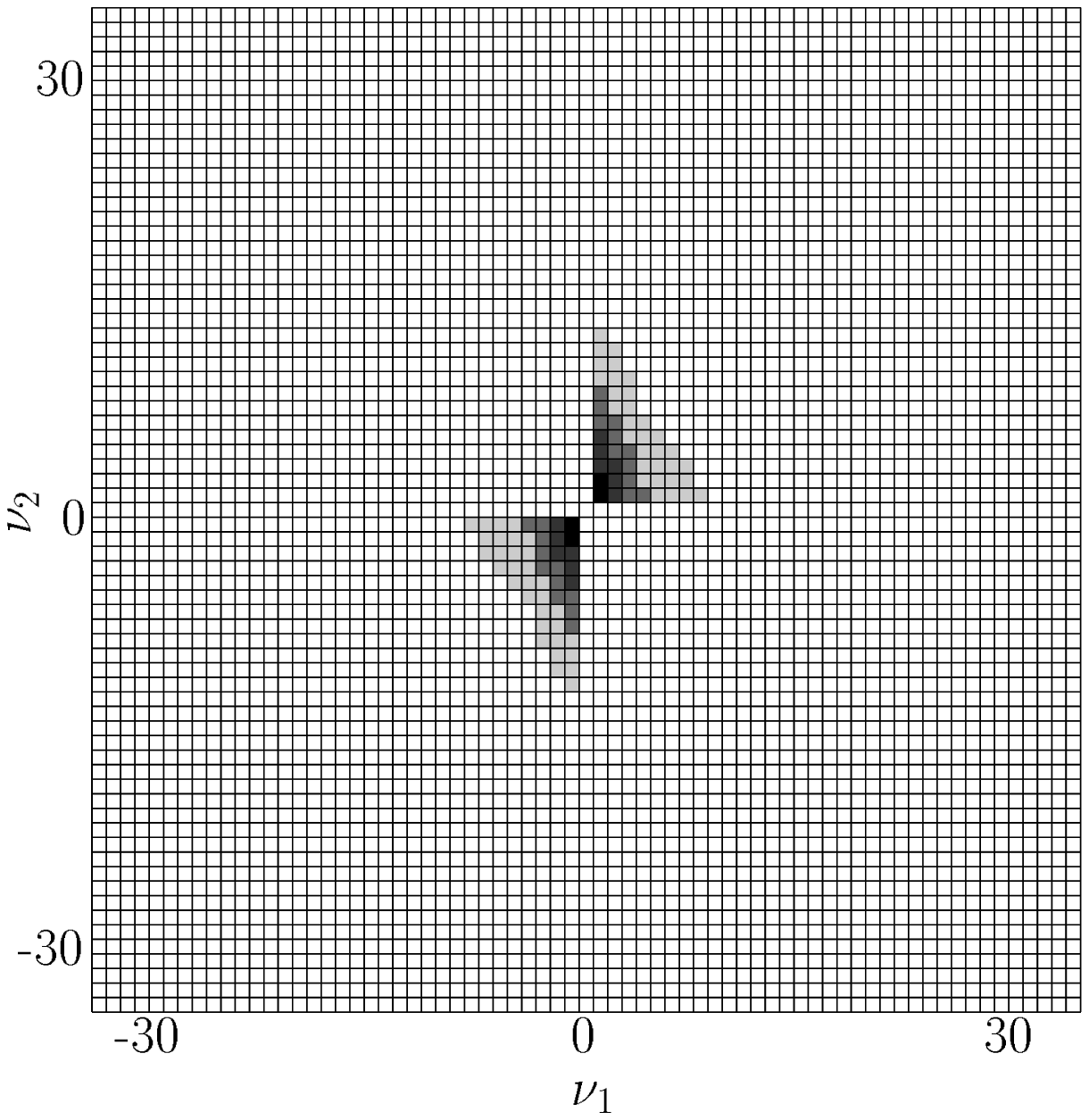,height=15cm,width=10cm}}
\normalsize 
\put(3,2){
FIG.\ 5. Weight of the Fourier coefficients of $g_*$:}
\put(3,1.6){White: $\, <10^{-10}$, grey levels: $[10^{-10},10^{-7}]$,}
\put(3,1.2){$[10^{-7},10^{-5}]$,$[10^{-5},10^{-3}]$,
$[10^{-3},5.10^{-2}]$, black: $>5.10^{-2}$.}
\end{picture}}
\end{figure}

\begin{figure}
\large
\unitlength=1cm
\centerline{
\begin{picture}(15,10)
\put(1.5,0){\psfig{figure=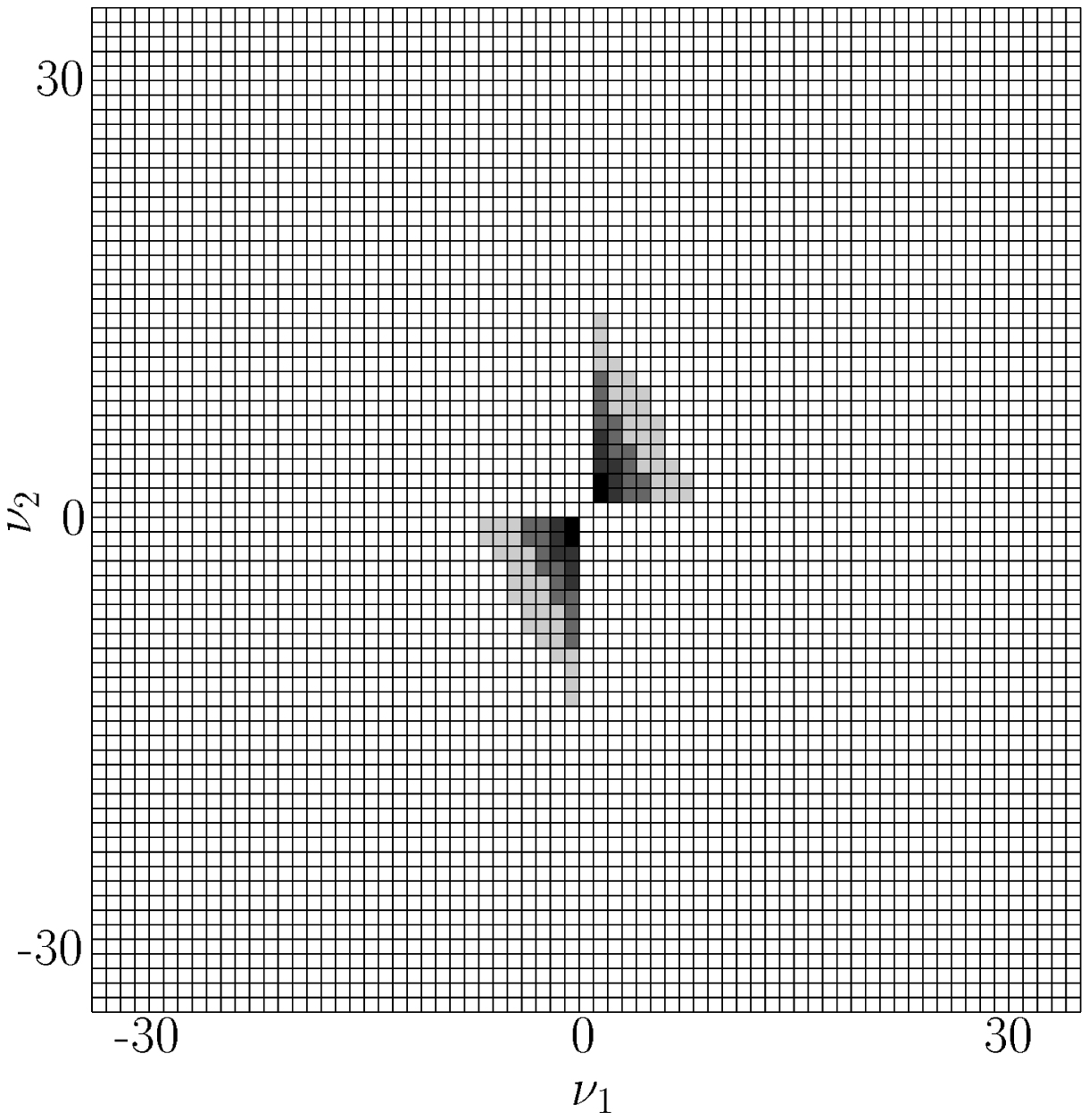,height=15cm,width=10cm}}
\normalsize 
\put(3,2){
FIG.\ 6. Weight of the Fourier coefficients of $f_*$:}
\put(3,1.6){White: $\, <10^{-10}$, grey levels: $[10^{-10},10^{-7}]$,}
\put(3,1.2){$[10^{-7},10^{-5}]$,$[10^{-5},10^{-3}]$,
$[10^{-3},5.10^{-2}]$, black: $>5.10^{-2}$.}
\end{picture}}
\end{figure}

\end{document}